\documentclass[prd,aps,floats,twocolumn,nofootinbib]{revtex4-1}
\usepackage{slashed}
\usepackage{mathtools}
\usepackage{amsfonts}
\usepackage{amssymb}
\usepackage{epsfig}
\usepackage{empheq}
\usepackage{enumitem}
\usepackage{amsmath} 
\usepackage{blkarray} 
\usepackage{multirow}
\usepackage{graphicx} 
\usepackage{float} 
\usepackage{xcolor}
\usepackage{amsmath,amssymb,mathtools}

\usepackage{bm}

\usepackage[polutonikogreek,english]{babel}
\usepackage[or]{teubner}

\usepackage{xcolor}

\begin{document}
\newcommand*\widefbox[1]{\fbox{\hspace{2em}#1\hspace{2em}}}
\newcommand{\m}[1]{\mathcal{#1}}
\newcommand{\nn}{\nonumber}
\newcommand{\ph}{\phantom}
\newcommand{\eps}{\epsilon}
\newcommand{\be}{\begin{equation}}
\newcommand{\ee}{\end{equation}}
\newcommand{\bea}{\begin{eqnarray}}
\newcommand{\eea}{\end{eqnarray}}
\newtheorem{conj}{Conjecture}

\newcommand{\plk}{\mathfrak{h}}


\title{Galactic dark matter halos: From anisotropic fluids in general relativity to Horava-Lifshitz gravity}
\date{}
\author{Paolo M Bassani}
\email{paolo.bassani22@imperial.ac.uk}
\affiliation{Abdus Salam Centre for Theoretical Physics, Imperial College London, Prince Consort Rd., London, SW7 2BZ, United Kingdom}

\begin{abstract}
We deform the GR Hamiltonian by adding an extra weight $+1$ density to the potential. We show that potential deformations of this type leave the (reduced) Dirac algebra unchanged and the modification is naturally reinterpreted as an effective anisotropic stress–energy contribution. While the fluid reproduces an isothermal-like mass scaling, its pressure anisotropy prevents it from giving flat rotation curves in this reduced phenomenological toy model.
We then turn to HL gravity where the absence of a local Hamiltonian constraint leaves a non-vanishing local Hamiltonian density, giving a controlled nonconservation law for the emergent dust component. Generalizing earlier results, we identify a restricted class of LTB backgrounds for which the HL source term yields a positive scaling dark matter density, consistent with ghost-freedom and with the continuous $\lambda \rightarrow 1$ limit, in which the effective dust amplitude vanishes together with the deviation from GR. The analysis is conditional on a prescribed background: obtaining a fully backreacted areal radius solution consistent with the HL field equations is left as a natural direction for future work.
\end{abstract}

\maketitle

\section{Introduction}

Galaxy rotation curves provide one of the cleanest dynamical indications that gravity and luminous matter, as modeled in standard General Relativity (GR) with baryons alone, do not account for observed galactic kinematics. Long-slit optical spectroscopy and 21-cm H I measurements show that the circular velocity of disk galaxies remains approximately constant out to radii well beyond the bright stellar disk, rather than declining in the Keplerian manner expected from the observed distribution of starlight and gas~\cite{Rubin:1980zd, Bosma:1981zz, Sofue:2001rv}. Within the standard paradigm, this “mass discrepancy” is attributed to a dominant, extended halo of non-luminous matter, commonly modeled as cold dark matter (CDM), whose gravitational potential can reproduce broadly flat rotation curves when combined with the baryonic components. This proposed solution postulates a particle nature for dark matter, as pointed out in~\cite{Sofue:2001rv, Salucci:2019dm, Wechsler:2018wpk}.
Despite its success as a phenomenological fit to galaxy rotation curves and its central role in the ΛCDM cosmological framework, the CDM interpretation is accompanied by persistent questions on galactic (and sub-galactic) scales~\cite{Bullock:2017xww}, and, crucially, by the absence, to date, of a confirmed laboratory detection of the underlying particle~\cite{PDG:2024rpp, Schumann:2019eaa}.

This motivates renewed scrutiny of the alternative logical possibility: that the rotation-curve anomaly reflects not an additional matter component in galaxies, but an incompleteness of the GR description, as argued in~\cite{Sanders:2002pf, Ferreira:2009mm, Bekenstein:2010mg}. In this view, one seeks minimal or controlled modifications of the gravitational sector—exemplified historically by MOND-like phenomenology~\cite{Milgrom:1983ca}—capable of reproducing the observed near-flat rotation curves while remaining consistent with established tests in the Solar System and with gravitational lensing and cosmological constraints.

In this work, we extend both possibilities by pursuing an unconventional link between dark matter in galactic halos and broken symmetries in GR and Horava-Lifshitz (HL)~\cite{HL} leading to violations of energy conservation. On the one hand, as pointed out in~\cite{Magueijo:2024mach, Magueijo:2024dmssr, Magueijo:2024stsb}, controlled violations of energy conservation (arising, for example, from evolving physical laws~\cite{Bassani:2025hmu, Magueijo:2023eco}) can produce dark matter like dust fluids. Inspired by these results, we modify the reduced GR theory by adding a new term to the potential. Although introduced on the geometric side, any potential deformation that preserves the (reduced) first-class structure is indistinguishable from an effective stress-energy, as argued in~\cite{Carroll:2000fy}. Our results confirm this, establishing a link between deformations in the potential and Dirac's hypersurface deformation algebra (HDA). Hence, within the reduced LTB framework, our GR-side modification is more properly interpreted as an effective anisotropic fluid than as a genuine modified-gravity theory with energy-conservation violations. At this phenomenological level it reproduces the expected density scaling, but still fails to support flat rotation curves.
On the other hand, HL theory reduces the full 4D space-time symmetry to a 3D spatial one, supplemented by a global time invariance. This reduced symmetry modifies the Hamiltonian constraint, producing an effective pressureless dust component whose conservation equation is violated. As pointed out in~\cite{shinji_review} and~\cite{Bassani:2025hl}, the associated source term can generate scaling profiles compatible with singular-isothermal-sphere (SIS) scaling and with dark-matter-like rotation-curve profiles in LTB spacetimes. Leveraging this result, we find a class of solutions leading to positive scaling dark matter energy density and flat rotational profiles, while setting strict constraint on the allowed parameters.

The two constructions answer a single question: within one and the same reduced LTB sector, which properties of the Hamiltonian constraint decide whether an effective $1/R^2$ source can support flat rotation curves? The GR-side deformation supplies the control case. Its failure is traceable to the feature that defines that case: the constraint is imposed pointwise and the HDA closes, which fixes the effective source through the unchanged momentum constraint and evolution equations. The minimal way to evade it is therefore to relinquish exactly that structure, which is what projectable HL gravity does, imposing the Hamiltonian constraint only globally. We use it here as a classical low-energy effective description, with $\lambda$ an infrared coupling constrained by present-day galactic kinematics and no input from the ultraviolet completion or from early-universe dynamics~\cite{shinji_review, Mukohyama:2009dm}.

This paper is therefore structured as follows: Section \ref{S2} reviews the LTB reduction of the GR Hamiltonian, introduces the $\alpha$-deformation, and briefly discusses the closure of the reduced constraint algebra. Furthermore, it presents a simple solution for the new field equations. Section \ref{S3} then develops the effective fluid interpretation, and studies rotation-curves in the reduced LTB test-tube. Finally, Section \ref{S5} generalizes the analysis to HL gravity, identifying the restricted class of backgrounds that yield a positive SIS-like dark matter scaling.

\section{LTB-Reduced Theory} \label{S2}

We study the effective minisuperspace reduction of GR and its possible extension with extra terms in the potential. To do so, we begin from the Einstein-Hilbert action expressed in canonical form ~\cite{Arnowitt:1959ah, Arnowitt:1962dg, Jha:2023hamadm}:
\begin{equation}
    S = \frac{1}{16 \pi G} \int_{t_1}^{t_2}{dt}\int_{\Sigma_t}{d^3 x} \: [\dot{h}_{ij}\pi^{ij} - N\mathcal{H}- N_i \mathcal{H}^i], \label{EH}
\end{equation}
where the Hamiltonian constraint $\mathcal{H}$ and the momentum constraint $\mathcal{H}_i$ are defined as:
\begin{align}
     \mathcal{H} &= -  \sqrt{h} \: {}^{(3)}R + \frac{1}{\sqrt{h}} \biggl(\pi_{ij} \pi^{ij}-\frac{\pi^2}{2}\biggl), \label{Hc}\\
     \mathcal{H}_i &= -2h_{ij} D_k \pi^{jk}\,. \label{Hi}
\end{align}
Here, $N(x, t)$ is the lapse function, $N_i (x,t)$ is the shift function, ${}^{(3)}R$ is the 3D intrinsic scalar curvature and $\pi_{ij}$ is the canonical momentum. We proceed by considering the LTB metric as presented in ~\cite{stephani, Enqvist:2007vb, Goncalves:2001rv}, a spherically symmetric dust solution with radial inhomogeneities, whose line element is:
\begin{equation}
    ds^2 = -dt^2 + X^2(r,t) dr^2 +A^2 (r,t) d\Omega^2, \label{ltb_metric}
\end{equation}
where $A(r, t)$ is the areal coordinate and $X(r, t)$ controls the radial proper distance. Because the lapse function is constant and the shift is zero, the LTB metric describes comoving, synchronous dust.
To express equation \eqref{Hc} and \eqref{Hi} in terms of LTB variables, we parametrize the canonical momenta $\pi^{ij}$~\cite{Vaz:2011cqsd} as:
\begin{equation}
    \pi^{ij} = diag\biggl(2G \sin \theta \frac{P_X}{X} , G\sin \theta \frac{P_A}{A}, \frac{G}{\sin \theta}\frac{P_A}{A}\biggl).
    \label{momenta}
\end{equation}
With this, the Hamiltonian density and the momentum density become:
\begin{align}
    \mathcal{H} &= -G \biggl(\frac{P_X P_A}{A} - \frac{X P_X^2}{2A^2}\biggl)
    +\frac{1}{G} \biggl[-\frac{X}{2}-\frac{(A^\prime)^2}{2X} + \biggl(\frac{AA^\prime}{X}\biggl)^\prime \biggl] \label{ham} \\
    \mathcal{H}_r &= 4 G \sin \theta (A^\prime P_A - X P_X^\prime), \label{mom}
\end{align}
and we have the LTB momenta:
\begin{equation}
    P_X = -\frac{A \dot{A}}{G}, \: \: P_A = -\frac{1}{G} (A\dot{X} + X\dot{A}). \label{canmom}
\end{equation}

\subsection{Modified Phenomenological Theory}

Until this point, we have followed the standard GR derivation. We now consider adding an extra term to the Hamiltonian's potential. For now, the extra term will play out as a geometric addition, but we will soon see how it can be repackaged into an effective anisotropic fluid component. Therefore, we heuristically introduce the new term $\alpha X(r, t)$ in the potential of equation \eqref{ham}, such that the new Hamiltonian density is:
\begin{align}
    \mathcal{H} = -G \biggl(\frac{P_X P_A}{A} - \frac{X P_X^2}{2A^2}\biggl) \nonumber\\
    +\frac{1}{G} \biggl[-\frac{X}{2}-\frac{(A^\prime)^2}{2X} + \biggl(\frac{AA^\prime}{X}\biggl)^\prime + \: \alpha X \biggl], \label{new_H}
\end{align}
where $\alpha$ is a dimensionless coupling constant. In the reduced variables every potential term in \eqref{ham} scales as $X$ does, so the extra term $\alpha X$ sits alongside $-X/2$ with $\alpha$ dimensionless, as consistency with the rest of the Hamiltonian requires.
Because the added term $H_\alpha[N]=(\alpha/G)\int d^3x\,N X$ is a scalar density of weight $+1$, contains no momenta, and involves no radial derivatives, it is immediate that it does not deform the hypersurface-deformation algebra. Concretely, its contribution to $\{H(N),H(M)\}$ is proportional to $\int d^3x\,NM(\cdot)$, which is symmetric under $N \leftrightarrow M$ and therefore cancels in the antisymmetrised bracket. Introducing the smeared constraints
\begin{align}
    H(N) &= \int d^3x\,N\mathcal{H},\\
    H_i(N^i) &= \int d^3x\,N^i\mathcal{H}_i,
\end{align}
the three brackets therefore retain their standard GR form \cite{Dirac, DiracCanadian, Thiemann, Bojowald:2011book},
\begin{align}
    \{H_i(N^i),H_j(M^j)\} &= H_i([N,M]^i),\label{hda1}\\
    \{H(N),H(M)\} &= H_r\!\left[h^{rr}(N\partial_r M - M\partial_r N)\right],\label{hda2}\\
    \{H_r(N^r),H(N)\} &= H(N^r\partial_r N), \label{hda3}
\end{align}
with the usual structure function $h^{rr}=1/X^2$. Hence the full set of constraints remains first class and realises the undeformed Dirac algebra. The $\alpha$-term modifies the Hamiltonian functional, but not the canonical realisation of spacetime diffeomorphisms, and will therefore be interpreted below as an effective anisotropic fluid contribution. The momentum constraint of this new effective theory is unchanged, such that, setting \eqref{mom} equal to zero and using equation \eqref{canmom} we obtain:
\begin{equation}
    \frac{\dot{A}^\prime}{A^\prime} = \frac{\dot{X}}{X} \Rightarrow X (r, t) = \frac{A^\prime}{\sqrt{1-k(r)}}, \label{momsoln}
\end{equation}
where $k(r)$ is an integration constant. Conversely, the Hamiltonian constraint is affected by the extra term. Varying \eqref{EH} with respect to the lapse function and using \eqref{canmom} into \eqref{new_H} we obtain the $G^0_0$ equation in vacuum:
\begin{equation}
    \frac{\dot{A}^2 + k(r)}{A^2} +\frac{2\dot{A}\dot{A}^\prime +k^\prime (r)}{A A^\prime} -\frac{2\alpha}{A^2} = 0.
\end{equation}
For model completeness, we add the standard GR pressureless dust fluid with energy density $\rho_m$. This is defined by adding to action \eqref{EH} a matter action $S_M$, such that we can define a matter Hamiltonian:
\begin{equation}
    H_m = \int d^3x\,N\,\mathcal H_m,
    \qquad
    \mathcal H_m = \sqrt{h}\,\rho_m.
\end{equation}
Then, the full Hamiltonian constraint becomes:
\begin{equation}
    \mathcal{H}_{tot} = \mathcal{H} + \mathcal{H}_m \approx 0,
\end{equation}
giving the $G^0_0$ Einstein equation with a dust matter source:
\begin{equation}
        \frac{\dot{A}^2 + k(r)}{A^2} +\frac{2\dot{A}\dot{A}^\prime +k^\prime (r)}{A A^\prime} -\frac{2\alpha}{A^2} = 8\pi G \rho_m. \label{H_cons}
\end{equation}
Including this matter source will help us model dust in galaxies and compare it with the effects and nature of the extra term in $\alpha$ added above.

Another modification to the standard GR theory arises in the $G^r_r$ equation, while the angular equations $G^\theta_\theta = G^\phi_\phi$ are unaffected. The equation of motion for $P_X$ becomes:
\begin{equation}
    \dot{P}_X = -G\frac{P_X^2}{2A^2} +\frac{1}{2G} -\frac{1}{G}\frac{(A^\prime)^2}{2X^2} -\frac{\alpha}{G},
\end{equation}
which, combined with equation \eqref{momsoln} and the time derivative of \eqref{canmom} gives:
\begin{equation}
    \dot{A}^2 + 2A \Ddot{A} + k(r) = 2 \alpha\label{grr}.
\end{equation}
This source implies a radial anisotropic negative pressure (tension) due to the extra term in $\mathcal{H}$, as we will see in the next section.
Then, the first integral of equation \eqref{grr} is:
\begin{equation}
    \dot{A}^2 = \frac{F_m(r)}{A} - k(r) \: +2\alpha = \frac{F_m (r) + F_{h}(r, t)}{A} - k(r) , \label{FI}
\end{equation}
where $F_m(r)$ is the usual Misner-Sharp mass associated with the geometric mass sourced by the dust and defined, via the Hamiltonian constraint, as:
\begin{equation}
    F^\prime_m (r)= 8 \pi G A^2 A^\prime \rho_m. \label{cons_m}
\end{equation}
Additionally, $F_h(r, t)$ is the geometric mass associated with the extra term in $\mathcal{H}$ defined as:
\begin{equation}
    F_h (r, t) := 2\alpha A. \label{F_halo}
\end{equation}

\subsection{A Simple Solution}
We now move to investigate specific solutions for this  model. The simplest solution for $A(r, t)$ comes from considering the outer region of a dust galaxy, where $\rho_m (R > R_b) = 0$, and hence $F_m = F_0 = const = 2G M_b$, with $M_b$ being the baryonic mass enclosed in $R_b$. Here, the gravitational mass felt by test particles is given both by the baryonic mass enclosed in the galaxy and the halo mass from $F_h$. Then, the dynamical equation to solve is a simpler case of the first integral \eqref{FI}:
\begin{equation}
    \dot{A}^2 = \frac{F_0}{A} - k + 2\alpha. \label{FI_halo}
\end{equation}
This has exactly the same form as the standard LTB equation in \cite{Enqvist:2007vb}, with the effective curvature being shifted because of the $\alpha$ constant. In what follows $k$ is taken constant on the shells under consideration, so the branches can be labelled by fixed values of $k$. Therefore, depending on the sign of $E := 2\alpha - k$, one can have hyperbolic, parabolic and elliptic solutions. In the elliptic solution $E < 0$, for small $A(r, t)$, the $F_0 /A$ term dominates, giving positive a positive dynamics $\dot{A}^2 > 0$. However, for a large radius, the constant term $E$ dominates, and we have $\dot{A}^2 < 0$. Therefore, this solution describes a re-collapsing dynamics not compatible with large extended halos, so we discard it.

Instead, the parabolic ($E = 0 \Leftrightarrow 2\alpha = k$) solution offers more scenarios. Since $2\alpha = k$, equation \eqref{FI_halo} reduces to the standard marginally bound LTB dust equation, with the shell's expansion rate decaying with radius. Further restrictions arise on the values of $k$: the $k = 0$ solutions also implies $\alpha = 0$, which is the trivial GR limit without any dark matter halo. The $k = -1$ solution makes the halo density in \eqref{halo_profile} negative, violating the energy conditions, therefore being unphysical. The $k = 1$ case leads to a singularity in the momentum constraint \eqref{momsoln}, so the only option left is $0 < k < 1$ with $\alpha \in (0, 1/2)$. Within this case the shells follow the same marginally bound dynamics as in standard GR LTB, with the anisotropic fluid superposed; the halo is not static, since $\rho_h \propto A^{-2}$ tracks the areal radius.

Finally, we consider the hyperbolic solution, with $E > 0$. The general parametric solution to equation \eqref{FI_halo} is:
\begin{equation}
    A(\eta) = \frac{F_0}{2(2\alpha -k)} (\cosh \eta -1), \label{hypersoln}
\end{equation}
from which we can distinguish two regimes: the baryon dominated regime and the halo dominated one. On the one hand, for earlier times we have that $F_0 /A >> E$, such that equation \eqref{hypersoln} reduces to the standard parabolic solution as above. This option is not of interest as it still describes the baryon dominated dynamics. On the other hand, for later times and $F_0 /A << E$ we find the asymptotic coasting solution:
\begin{equation}
    A(r, t) = \sqrt{|2\alpha - k|} (t-t_0(r)),
\end{equation}
where $t_0(r)$ is the shell bang time.

\section{Fluid interpretation and rotation curves} \label{S3}
As mentioned before, one can interpret the extra geometric term in \eqref{H_cons} as a new fluid  by moving it to the matter sector as highlighted in~\cite{Carroll:2000fy}, such that \eqref{H_cons} now reads:
\begin{equation}
    \frac{\dot{A}^2 + k(r)}{A^2} +\frac{2\dot{A}\dot{A}^\prime +k^\prime (r)}{A A^\prime} = 8\pi G \rho_m + \frac{2\alpha}{A^2}.
\end{equation}
Then, comparing this with the Hamiltonian constraint of standard GR with matter plus an additional effective source $\rho_h$:
\begin{equation}
    \frac{\dot{A}^2 + k(r)}{A^2} +\frac{2\dot{A}\dot{A}^\prime +k^\prime (r)}{A A^\prime} = 8\pi G (\rho_m + \rho_h),
\end{equation}
we can immediately make the identification:
\begin{equation}
    \rho_{h} := \frac{\alpha}{4 \pi G}\frac{1}{A^2}.
    \label{halo_profile}
\end{equation}
Similarly, using the modified evolution equation \eqref{grr} and comparing it with the standard $G^r_r$ equation with pressure in LTB:
\begin{equation}
    \dot{A}^2 + 2A \Ddot{A} + k(r) = -8\pi G p_r A^2,
\end{equation}
we see that the fluid's radial pressure is:
\begin{equation}
    p_r = -\frac{\alpha}{4 \pi G}\frac{1}{A^2}.
\end{equation}

Furthermore, since the angular equations $G^{\theta}_\theta = G^{\phi}_\phi$ are unaffected by the $\alpha$ term, there is no tangential pressure.
At this stage it is important to distinguish radial momentum flux from radial pressure in the canonical language. In the ADM split \cite{Jha:2023hamadm}, an effective source is characterized by the energy density
\begin{equation}
E = T_{\mu\nu}n^\mu n^\nu,
\end{equation}
the momentum density (or energy flux)
\begin{equation}
J_i = - h_i{}^\mu n^\nu T_{\mu\nu},
\end{equation}
and the spatial stress tensor
\begin{equation}
S_{ij} = h_i{}^\mu h_j{}^\nu T_{\mu\nu}.
\end{equation}
The scalar constraint determines $E$, the diffeomorphism constraint determines $J_i$, while the evolution equations determine $S_{ij}$. Therefore, since the $\alpha$-term leaves $\mathcal H_r$ unchanged, the effective source carries no additional radial momentum density,
\begin{equation}
J_r=0.
\end{equation}
This does not forbid a non-vanishing radial pressure: $p_r=S^r{}_r$ is a spatial stress, not a flux, and is instead reconstructed from the modified evolution equation for $P_X$ (equivalently, from the $G^r{}_r$ equation).
These findings imply an anisotropic equation of state, with radial and tangential components:
\begin{align}
    p_r &= -\rho_h \Rightarrow w_r = -1 \\
    p_t &= 0 \Rightarrow w_t = 0,
\end{align}
and stress-energy momentum tensor:
\begin{equation}
    T^{\mu}_{\nu , h} = diag (-\rho_h , -\rho_h, 0, 0).
\end{equation}
This is formally the same algebraic structure encountered in classical GR descriptions of global monopole exteriors~\cite{Barriola:1989hx} and of spherically symmetric string-cloud/string-hedgehog sources~\cite{Letelier:1979ej}. We stress, however, that our observation is restricted to this stress–energy-level correspondence within the reduced model: it does not claim the same microphysical origin, topological content, or full spacetime solution as genuine monopole or string configurations in the underlying (unreduced) theory.
We can then derive the continuity equation for this new fluid. Considering the conservation of $T^{\mu}_{\nu, h}$ and the pressure we obtain:
\begin{equation}
    \dot{\rho}_h + 2 \frac{\dot{A}}{A} \rho_h = 0,
\end{equation}
which, expectedly, has \eqref{halo_profile} as solution, once the integration constant is suitable fixed. The radial component of the same conservation law, $\partial_r p_r + 2(p_r - p_t)A^\prime /A = 0$, is satisfied identically by $p_r = -\alpha/(4\pi G A^2)$ and $p_t = 0$, so the effective-fluid interpretation is consistent in both components. This effective halo fluid is strictly conserved: since the added term is a spatial scalar density of weight $+1$ and leaves the (reduced) hypersurface-deformation algebra unchanged, it does not break the Hamiltonian constraint symmetry, and no source term can appear in the continuity equation. Conversely, a genuine source term could arise from modifications such as explicit time dependence (e.g. $\alpha(t)$), preferred-foliation terms involving lapse derivatives $a_i = D_i \ln N$ (in a style very similar to HL), non–weight-$+1$ additions, or nonlinear lapse-dependent potentials like $\sqrt{h} F(N)$. We do not pursue these possibilities here and leave their phenomenological explorations for future work.

Finally, we can consider the energy conditions~\cite{Curiel:2017primerEC} satisfied by this fluid. The null energy condition is satisfied both in the radial and tangential directions, with saturation in the former. The weak energy condition is satisfied in the same way as the null one. The dominant energy condition is:
\begin{equation}
    \rho_h \geq 0 , \alpha > 0,
\end{equation}
and
\begin{equation}
    \rho_h = |p_r|,
\end{equation}
again with saturation in the radial direction. The strong energy condition, once again, is satisfied with radial saturation. Saturating all the energy conditions means the fluid sits exactly on the boundary between ordinary and exotic matter. However, physically, it can still support all the standard GR theorems, while behaving in a very fine-tuned way.

\subsection{Rotation Curve Test}
The previous subsection established the dynamics without requiring any fluid language; here we simply reinterpret the extra geometric contribution $F_h=2\alpha A$ from equation \eqref{F_halo} in terms of the effective source $(\rho_h,p_r,p_t)$ reconstructed from the constraint and evolution equations. This allows us to test whether the same canonical solution, viewed as an effective halo, leads to the phenomenology of approximately flat galactic rotation curves. We stress, however, that this should be understood only as a reduced phenomenological test within the spherically symmetric LTB sector: we are not claiming that the present $\alpha$-deformation defines a complete four-dimensional modified-gravity theory, nor a full GR model with a specified microscopic matter source for realistic galaxies. Rather, the point is to ask a more limited question: if such a deformation is introduced in the reduced canonical theory and then reinterpreted as an effective source, does the resulting source even belong to the class capable of supporting approximately flat rotation curves?

In the quasi-static regime, one may treat
$A(r, t) \approx A(r) \equiv R$ as an areal radius and interpret the additional geometric term as an effective halo component with
\begin{equation}
    M(R) = 4\pi \int_0^R{\rho_h(r) r^2 dr} = \frac{\alpha}{G}R.
\end{equation}
Thus the modification in the Hamiltonian reproduces the characteristic isothermal scaling that is usually associated with flat rotation curves, as pointed out in \cite{Treu:2010SLreview}. However, this very same construction results in an anisotropic energy-momentum tensor. Since pressure contributes to the relativistic gravitational field \cite{Komar:1963pde}, the usual Newtonian circular velocity is not applicable: for this component the “active” mass combination $\rho_h  + p_r + 2p_t$ vanishes identically, so the linear growth of Misner–Sharp mass does not translate into genuinely flat circular velocities. In this sense, the present modification generates the right radial scaling but in the wrong universality class, behaving more like a defect/string-type anisotropic medium than like pressureless dark matter.
Read in this way, the rotation-curve discussion is not meant to promote the $\alpha$-term as a realistic halo model, but rather to isolate the phenomenological consequence of the reduced canonical construction itself. In other words, the spherical reduction is being used here as a controlled diagnostic setting: if the mechanism already fails to produce the correct effective stress structure in this simplest sector, then it cannot by itself be regarded as a viable explanation of galactic flat curves. The result is therefore best understood as a no-go statement for this class of potential deformations within the reduced GR framework, rather than as a completed model of galaxy dynamics.

This outcome is structural within the present canonical setup: the added term alters only the potential while leaving the kinetic part of the Hamiltonian, and therefore the HDA, unchanged, so the theory remains in the GR covariance class and the effective source is forced into the specific anisotropic form above. The same holds for any deformation that stays in that covariance class: while the constraint is imposed pointwise and the HDA closes, the effective source is conserved and its stress structure is fixed by the unchanged $\mathcal{H}_r$ and evolution equations. Evading the obstruction therefore requires relinquishing the pointwise Hamiltonian constraint itself, the defining feature of the projectable theory. It is therefore natural to contrast this mechanism with Hořava–Lifshitz constructions \cite{HL}, where an effective dust component arises not from a conserved anisotropic stress but from a controlled violation of local energy conservation along the preferred foliation \cite{Bassani:2025hl} and \cite{shinji_review}. In the next section we show explicitly how the HL source term can generate a dust density with $\rho_h \propto 1/A^2$. We then explain why the same scaling corresponds there to isotropic pressureless matter, hence capable of supporting flat rotation curves. Read together, the two sections form a matched pair: one reduced LTB sector, one structural difference.

\section{Evading the obstruction: projectable Ho\v{r}ava--Lifshitz gravity} \label{S5}

We now remove the structure identified above and keep everything else fixed. The infrared limit of the projectable theory is not GR alone: it is GR together with one extra scalar degree of freedom, which at low energies behaves as pressureless dust \cite{Mukohyama:2009dm, shinji_review}. GR is recovered continuously as $\lambda \rightarrow 1$, and the amplitude of that dust vanishes in the same limit; the dust is the $O(\lambda-1)$ deviation from GR, not a component that survives it. What follows is a statement about this infrared structure evaluated today: the source term of the nonconservation law acts on a present-day quasi-static galactic background and drives the density there, so no relic abundance and no propagation through nonlinear cosmological history enters the argument.

As already pointed out in \cite{HL}, projectable Ho\v{r}ava--Lifshitz gravity
differs from GR by reducing the symmetry from full spacetime diffeomorphism
invariance to foliation-preserving diffeomorphisms,
\begin{equation}
    t\rightarrow t'(t),\qquad x^i\rightarrow x^{\prime i}(t,x).
\end{equation}
The lapse is therefore projectable, \(N=N(t)\), while the shift remains local.
As a consequence, the Hamiltonian constraint is imposed only globally,
\begin{equation}
    H_0=\int_{\Sigma_t}d^3x\,\mathcal H(x)\approx0,
\end{equation}
whereas the local momentum constraint remains
\begin{equation}
    \mathcal H_i(x)\approx0.
\end{equation}
Thus the local Hamiltonian density \(\mathcal H(x)\) need not vanish
pointwise. This leftover local Hamiltonian density is the canonical expression
of the extra scalar degree of freedom of the projectable theory and, at low
energies, can be equivalently described as a pressureless dust component.

This equivalence is implemented by assigning the non-vanishing local
Hamiltonian to an effective matter contribution,
\begin{equation}
    \mathcal H_{\rm DM}:=-\mathcal H,
\end{equation}
so that the enlarged local Hamiltonian density satisfies
\begin{equation}
    \overline{\mathcal H}
    =
    \mathcal H+\mathcal H_{\rm DM}
    \approx0 .
    \label{HL_repackaging}
\end{equation}
Equation \eqref{HL_repackaging} is not a restoration of the non-projectable
HL theory. Rather, it is a local effective description of the same projectable
configuration, in which the non-zero gravitational Hamiltonian density is
carried by an effective dust sector. The preferred foliation remains geodesic
and the lapse remains projectable on shell, but the effective description allows
one to identify the stress tensor associated with the leftover Hamiltonian.

For the infrared HL action we take the Hamiltonian and momentum densities to be:
\begin{align}
     \mathcal{H} &=
     \frac{2}{M_{\rm Pl}^2\sqrt{h}}
     \biggl[
        \pi_{ij}\pi^{ij}
        -
        \biggl(
            \frac{\lambda}{3\lambda -1}
        \biggr)\pi^2
     \biggr]
     -\sqrt{h}\,{}^{(3)}R ,
     \label{HL_H}
     \\
     \mathcal{H}_i &= -2h_{ij}D_k\pi^{jk},
\end{align}
where \(M_{\rm Pl}\) is the reduced Planck mass and \(\lambda\) is the
dimensionless HL kinetic coupling. We work in units in which the low-energy
gravitational light speed \(c_g\), carried by the curvature term, is set to unity, and we
restore it explicitly in the nonconservation law below. With the smeared functionals
\begin{equation}
    H[N]=\int d^3x\,N\mathcal H,
    \qquad
    H_i[N^i]=\int d^3x\,N^i\mathcal H_i,
\end{equation}
the relevant Poisson brackets are:
\begin{align}
    \{H_i[N^i],H_j[M^j]\}
    &=
    H_i\!\left[[N,M]^i\right],
    \\
    \{H_i[N^i],H[M]\}
    &=
    H[N^i\partial_iM],
    \\
    \{H[N],H[M]\}
    &=
    C_i\!\left[
        h^{ij}
        \left(
            N\partial_jM-M\partial_jN
        \right)
    \right],
    \label{HL_HH_bracket}
\end{align}
where
\begin{equation}
    C_i
    =
    \mathcal H_i+2\hat\lambda D_i\pi,
    \qquad
    \hat\lambda
    =
    \frac{\lambda-1}{3\lambda-1}.
    \label{HL_Ci}
\end{equation}
The last equation is the point at which the HL kinetic term differs from the
GR one. In GR, or in the limit \(\lambda=1\), \(C_i\) reduces to the momentum
constraint \(\mathcal H_i\). In projectable HL the additional
\(D_i\pi\) term remains and controls the propagation of the leftover local
Hamiltonian density.
Therefore, the role of equation \eqref{HL_HH_bracket} is to evolve the non-vanishing
local density \(\mathcal H(x)\), not to impose \(\mathcal H(x)\approx0\) as a
constraint of the bare projectable theory. The full Hamiltonian generating the
projectable dynamics is:
\begin{equation}
    H_{\rm HL}
    =
    \int d^3x
    \left[
        N(t)\mathcal H
        +
        N^i(t,x)\mathcal H_i
    \right].
\end{equation}
Using \eqref{HL_HH_bracket} and \eqref{HL_Ci}, the local Hamiltonian density
propagates as:
\begin{equation}
    \dot{\mathcal H}
    =
    \{\mathcal H,H_{\rm HL}\}
    =
    \partial_i(N^i\mathcal H)
    +
    D_i(NC^i).
    \label{HL_Hdot_general}
\end{equation}
Since the lapse is projectable, \(D_iN=0\). Moreover, the momentum constraint
is a genuine local constraint and is preserved on the preferred geodesic
foliation. Evaluating \eqref{HL_Hdot_general} on the surface
\(\mathcal H_i=0\), one obtains
\begin{equation}
    \dot{\mathcal H}
    =
    \partial_i(N^i\mathcal H)
    +
    2\hat\lambda N D^2\pi .
    \label{HL_Hdot}
\end{equation}
This is the local evolution equation for the leftover Hamiltonian density. It
is not the preservation equation of a local Hamiltonian constraint; the local
Hamiltonian constraint has precisely been lost in the projectable theory.

Since \(\mathcal H_{\rm DM}=-\mathcal H\), equation \eqref{HL_Hdot} gives
\begin{equation}
    \dot{\mathcal H}_{\rm DM}
    =
    \partial_i(N^i\mathcal H_{\rm DM})
    -
    2\hat\lambda N D^2\pi .
    \label{HL_HDMdot}
\end{equation}
The effective component has Hamiltonian density \(\mathcal H_{\rm DM}\), no
momentum density, and no spatial stresses. The absence of a momentum density
follows from \(\mathcal H_i\) remaining an unmodified local constraint; the absence of spatial
stresses is the standard low-energy behaviour of the extra scalar mode of the projectable
theory, which propagates as pressureless dust \cite{Mukohyama:2009dm}. Its stress tensor is therefore:
\begin{equation}
    T^{\mu\nu}_{\rm DM}
    =
    \rho_{\rm DM}n^\mu n^\nu,
    \qquad
    \rho_{\rm DM}
    =
    \frac{\mathcal H_{\rm DM}}{\sqrt h}.
    \label{HL_dust_tensor}
\end{equation}
The non-trivial component is the projection along \(n^\mu\):
\begin{equation}
    -n_\mu\nabla_\nu T_{\rm DM}^{\mu\nu}
    =
    \nabla_\mu(\rho_{\rm DM}n^\mu)
    =
    \frac{1}{N\sqrt h}
    \left[
        \dot{\mathcal H}_{\rm DM}
        -
        \partial_i(N^i\mathcal H_{\rm DM})
    \right].
    \label{HL_divergence}
\end{equation}
Substituting \eqref{HL_HDMdot} into \eqref{HL_divergence} gives
\begin{equation}
    -n_\mu\nabla_\nu T_{\rm DM}^{\mu\nu}
    =
    -\frac{2c_g^2\hat\lambda}{\sqrt h}D^2\pi .
    \label{HL_noncons_pi}
\end{equation}
Finally, using the trace of the canonical momentum:
\begin{equation}
    \pi
    =
    \frac{M_{\rm Pl}^2}{2}
    \sqrt h\,(1-3\lambda)K,
\end{equation}
where \(K\) is the extrinsic curvature scalar, this becomes:
\begin{equation}
    -n_\mu\nabla_\nu T_{\rm DM}^{\mu\nu}
    =
    c_g^2M_{\rm Pl}^2(\lambda-1)D^2K .
    \label{HL_noncons_K}
\end{equation}
Thus the effective dust component associated with the leftover Hamiltonian is
conserved in the GR limit \(\lambda=1\), but is sourced in projectable HL by the
spatial Laplacian of the extrinsic curvature. The source is not due to imposing
a local Hamiltonian constraint in the projectable theory; it is due to the HL
propagation of the non-vanishing local Hamiltonian density that has been
repackaged as an effective dust Hamiltonian.

We now specialize the non-conservation equation \eqref{HL_noncons_K} to a
spherically symmetric LTB-type geometry, whose metric is given in \eqref{ltb_metric}. In this metric, the trace of the extrinsic curvature is:
\begin{equation}
    K
    =
    \nabla_\mu n^\mu
    =
    \frac{\dot A'}{A'}
    +
    2\frac{\dot A}{A}.
    \label{K_LTB}
\end{equation}
Then, using:
\begin{equation}
    -n_\mu\nabla_\nu T_{\rm DM}^{\mu\nu}
    =
    \dot\rho_{\rm DM}
    +
    K\rho_{\rm DM},
\end{equation}
the HL non-conservation law becomes:
\begin{equation}
    \dot{\rho}_{\rm DM}
    +
    \rho_{\rm DM}
    \left[
        \frac{\dot A'}{A'}
        +
        2\frac{\dot A}{A}
    \right]
    =
    c_g^2M_{\rm Pl}^2(\lambda-1)D^2K ,
    \label{rho_HL_LTB}
\end{equation}

We now want to study how the source term in equation \eqref{rho_HL_LTB} can generate dark matter-like scaling profiles in galaxies. In the quasi-static pressureless limit, flat rotation curves correspond to the
singular-isothermal-sphere scaling \(\rho(R)\propto R^{-2}\)
\cite{Treu:2010SLreview}. Since the areal radius on a fixed time slice is
\(R=A(t_\star,r)\), the HL source must drive a contribution scaling as:
\begin{equation}
    \rho_{\rm DM}^{\rm driven}\propto \frac{1}{A^2}.
\end{equation}
For simplicity, we first ignore the overall constant and substitute
\(\rho_{\rm DM}^{\rm driven}=A^{-2}\) into the left-hand side of
\eqref{rho_HL_LTB}. This gives:
\begin{equation}
    \dot{\rho}_{\rm DM}^{\rm driven}
    +
    \rho_{\rm DM}^{\rm driven}
    \left[
        \frac{\dot A'}{A'}
        +
        2\frac{\dot A}{A}
    \right]
    =
    \frac{\dot A'}{A'A^2}. \label{source}
\end{equation}
Thus, in order to produce the SIS-like scaling, the HL source term must have
the same dependence on the LTB variables,
\begin{equation}
    D^2K \propto \frac{\dot A'}{A'A^2}.
    \label{source_required}
\end{equation}
A generic LTB background does not necessarily satisfy this condition: in
general, \(D^2K\) can generate a different radial profile, and hence a different
effective dust scaling. We therefore restrict to a simple flat LTB test
background for which this structure can be checked explicitly,
\begin{equation}
    A(t,r)=A_0h(s),
    \qquad
    s=r-r_0-t,
    \label{HL_ansatz}
\end{equation}
for which
\begin{equation}
    \dot A=-A',
    \qquad
    \dot A'=-A''.
\end{equation}
The background is flat in the sense of \eqref{momsoln} with \(k=0\), so that
\(X=A'\); this is the form used in evaluating \(D^2K\) below.
Hence, using equation \eqref{K_LTB} and factoring out the source term in equation \eqref{source} that would give flat rotation curves, we obtain:
\begin{equation}
    D^2K
    =
    \Psi(h(s))\,
    \frac{\dot A'}{A'A^2},
    \label{D2K_factorized}
\end{equation}
where derivatives of \(h\) are with respect to \(s\), and
\begin{align}
    \Psi (h(s))
    = \frac{h^2 h^{(4)}}{(h^\prime)^2 h^{\prime \prime}} +4\frac{h h^{\prime \prime \prime}}{h^\prime h^{\prime \prime}} \nonumber\\
    -4\frac{h h^{\prime \prime}}{(h^\prime)^2}-4\frac{h^2 h^{\prime \prime \prime}}{(h^\prime)^3}+3\frac{h^2 (h^{\prime \prime})^2}{(h^\prime)^4}.\label{psi}
\end{align}
Therefore, on this restricted background, the HL non-conservation equation becomes:
\begin{equation}
    \dot{\rho}_{\rm DM}
    +
    \rho_{\rm DM}
    \left[
        \frac{\dot A'}{A'}
        +
        2\frac{\dot A}{A}
    \right]
    =
    c_g^2M_{\rm Pl}^2(\lambda-1)
    \Psi(h(s))\,
    \frac{\dot A'}{A'A^2}.
    \label{cons_HL}
\end{equation}

To obtain a genuine \(1/A^2\) driven contribution, \(\Psi(h(s))\) must clearly be constant. Otherwise the source would produce a more general radial profile
rather than the SIS-like scaling. A simple non-trivial choice is the power-law
profile
\begin{equation}
    h(s)=|s|^p,
\end{equation}
for which
\begin{equation}
    \Psi(h(s))=\frac{1-3p}{p^2}.
\end{equation}
By contrast, an exponential profile gives \(\Psi(h(s))=0\), so the HL source
vanishes and no driven dust component is produced. Hence the non-trivial case
of interest is the power-law ansatz. Hence, for the \(h(s)=|s|^p\) ansatz, equation \eqref{cons_HL} becomes:
\begin{equation}
    \dot{\rho}_{\rm DM}
    +
    \rho_{\rm DM}
    \left[
        \frac{\dot A'}{A'}
        +
        2\frac{\dot A}{A}
    \right]
    =
    c_g^2M_{\rm Pl}^2(\lambda-1)
    \frac{1-3p}{p^2}
    \frac{\dot A'}{A'A^2},
    \label{rho_powerlaw_eq}
\end{equation}
which can be straightforwardly solved to:
\begin{equation}
    \rho_{\rm DM}(t,r)
    =
    \frac{\mathcal C(r)}{A'A^2}
    +
    c_g^2M_{\rm Pl}^2(\lambda-1)
    \frac{1-3p}{p^2}
    \frac{1}{A^2},
    \label{hl_soln}
\end{equation}
where $\mathcal{C}(r)$ is an integration constant. The first term is the homogeneous, freely propagating dust contribution. The
second term is the driven dark matter contribution produced by the HL non-conservation
source.

Several restrictions follow immediately. The cases \(p=0\) and \(p=1\) are
degenerate in the factorized expression \eqref{psi}, while \(p=1/3\) makes the
driven source vanish. Moreover, the ghost-free branch continuously connected
to GR requires
\begin{equation}
    \lambda>1.
\end{equation}
On this branch, positivity of the driven density in \eqref{hl_soln} requires
\begin{equation}
    1-3p>0.
\end{equation}
Taking \(p>0\), the parameter range selected by positivity and infrared
viability is therefore
\begin{equation}
    \lambda>1,
    \qquad
    0<p<\frac13.
    \label{constraint}
\end{equation}
In the contracting branch, \(s=r-r_0-t\rightarrow0^+\), the driven term
dominates over the homogeneous term whenever \(p<1\). This condition is
automatically satisfied in \eqref{constraint}. Outside this branch, the
homogeneous contribution generally remains present unless the initial data are
chosen so that \(\mathcal C(r)\) is negligible.

Finally, we show how the driven term leads to flat rotation curves in this toy
model. On a quasi-static time slice \(t=t_\star\), define the areal radius:
\begin{equation}
    R:=A(t_\star,r),
\end{equation}
such that the driven density becomes:
\begin{equation}
    \rho_{\rm DM}^{\rm driven}(R)
    =
    c_g^2M_{\rm Pl}^2(\lambda-1)
    \frac{1-3p}{p^2}
    \frac{1}{R^2}.
    \label{driven}
\end{equation}
Assuming that the infrared weak-field limit is governed by the usual Poisson
equation with the locally measured Newton constant \(G_N\),
\begin{equation}
    \nabla^2\Phi=4\pi G_N\rho,
\end{equation}
and using spherical symmetry, one has:
\begin{equation}
    \frac{1}{R^2}
    \frac{d}{dR}
    \left(
        R^2\Phi'(R)
    \right)
    =
    4\pi G_N
    \rho_{\rm DM}^{\rm driven}(R).
\end{equation}
For the driven solution in equation \eqref{driven} this integrates to:
\begin{equation}
    \Phi'(R)
    =
    \frac{C_1}{R^2}
    + c_g^2M_{\rm Pl}^2(\lambda-1)
    \frac{1-3p}{p^2}
    \frac{4\pi G_N}{R}.
\end{equation}
The circular velocity is therefore:
\begin{equation}
    v_c^2
    =
    R\Phi'(R)
    =
    4\pi G_Nc_g^2M_{\rm Pl}^2(\lambda-1)
    \frac{1-3p}{p^2}
    +
    \frac{C_1}{R}.
\end{equation}
The second term decays at large radius, leaving the asymptotic value
\begin{equation}
    \frac{v_{\rm flat}^2}{c_g^2}
    =
    4\pi G_NM_{\rm Pl}^2
    (\lambda-1)
    \frac{1-3p}{p^2}.
\end{equation}
Using the reduced Planck mass convention \(M_{\rm Pl}^{-2}=8\pi G_N\), this
becomes
\begin{equation}
    \frac{v_{\rm flat}^2}{c_g^2}
    =
    \frac12
    (\lambda-1)
    \frac{1-3p}{p^2}.
    \label{vflat_HL}
\end{equation}

For a typical spiral galaxy like the Milky Way, $v_{\rm flat} \sim 200\,\mathrm{km/s}$ \cite{Sofue:2017rm, Persic:1995ru}. Then, given the constraints in \eqref{constraint}, we find that for a generic $p$ not too close to $1/3$ (for example, $p = [0.05, 0.3]$), $\lambda$ has to be extremely close to one, specifically $\lambda - 1 \sim [10^{-9}, 10^{-6}]$. Otherwise, if $\lambda$ was to significantly deviate from one, then $p$ would need to be extremely close to $1/3$. In other words, large deviations of $\lambda$ from one are incompatible with usual galactic $v_{\rm flat}$ unless $p$ is tuned extraordinarily close to $1/3$.

In this sense, flat curve phenomenology by itself does not provide a clean discriminator between HL gravity and GR at the level of the present construction. Taking observed galactic rotation curves at face value forces a stark alternative: either the infrared theory lies in a near-GR corner of parameter space, with $\lambda$ so close to unity that any HL-specific deviation is effectively undetectable in the quasi-static regime; or one permits $\lambda$ to depart appreciably from unity by pushing $p$ extremely close to $1/3$, in which case the same limit that relaxes the constraint on $\lambda$ simultaneously suppresses the HL-driven source term—rendering it parametrically tiny, and in the strict $p \rightarrow 1/3$ limit effectively zero. The latter option is therefore difficult to reconcile with the notion that the HL-induced contribution could dominate present-day galactic rotation curves, since the mechanism that allows sizeable deviations from GR also removes the very source responsible for the would-be effective dark component. The first branch of this alternative is not an inconsistency: the driven amplitude in \eqref{vflat_HL} is proportional to $(\lambda-1)$, so it vanishes in the same limit in which GR is recovered, and a small $\lambda-1$ measures the infrared departure from GR rather than signalling that a surviving dark component contradicts that recovery. The degeneracy with $p$ remains a genuine limitation.

Within these limitations, the analysis still identifies a regime in which the HL nonconservation law yields a positive $1/R^2$-type scaling for the effective dust density, improving on the sign issue highlighted in \cite{Bassani:2025hl}. At the same time, the result should be read as a conditional, “passive” statement: the power-law ansatz $A(r, t) \propto |s|^p $ is imposed to extract the scaling implied by the HL source term on a prescribed LTB background, not derived as a solution of the full HL field equations. Consequently, it does not address whether the assumed geometry remains consistent once backreaction is included, nor does it incorporate ordinary baryonic/dust matter that would inevitably contribute in realistic galaxies. This is directly analogous to the “test-tube” strategy employed in Section V of \cite{Bassani:2025hl}, where a fixed Schwarzschild/Lemaître background is adopted precisely to isolate the scaling generated by the HL term in a controlled limit. Finally, it is worth emphasizing that the present discussion is aimed at late-time, quasi-static galactic phenomenology: different considerations can plausibly apply in genuinely time-dependent settings such as the early Universe, where the relevant scales, background dynamics, and dominant matter content need not mirror the galactic regime, and where HL effects are often expected to be more pronounced.

\section{Conclusion}

In this work we examined, within a spherically symmetric LTB setting, whether “dark-matter-like” phenomenology can arise either from controlled deformations of the GR Hamiltonian or from the symmetry reduction characteristic of Hořava–Lifshitz gravity. The guiding motivation was the long-standing rotation-curve anomaly and the observation that, in a number of contexts, effective dust components can be generated by departures from standard conservation laws or by modified constraint structures.
On the GR side, we introduced an additional term in the Hamiltonian potential within the LTB reduction and followed its consequences through the field equations and the hypersurface-deformation algebra. We obtained two central results. First, deforming only the potential by adding a weight-one scalar density—such as the term considered here—does not produce the Hořava–Lifshitz mechanism: the constraint algebra closes exactly as in GR, so the theory remains in the same covariance class and the dust continuity equation is not genuinely violated. Achieving true nonconservation would require introducing terms that are not invariant under the underlying symmetry (for instance a time-dependent coupling $\alpha(t)$, or more generally deformations that fail to behave as scalar densities under the canonical gauge transformations), which we do not pursue here and leave for future work. Second, the deformation can be consistently repackaged as an effective anisotropic fluid with radial tension, which reproduces the desired isothermal-like scaling at the level of the continuity equation and mass profile; nevertheless, the same anisotropy obstructs the phenomenology of interest, since the resulting active gravitational mass does not behave like pressureless matter and therefore does not support genuinely flat circular-velocity curves. Furthermore, this analysis should be read as a reduced-model obstruction rather than as a realistic halo construction. The spherically symmetric LTB minisuperspace is used here as a controlled canonical setting in which one can determine unambiguously what kind of effective source is generated by a potential deformation that preserves the reduced Dirac algebra. The result is a no-go statement: while the induced source reproduces the familiar $1/R^2$ scaling at the level of density and Misner--Sharp mass, its unavoidable anisotropic stress makes it inequivalent to pressureless halo matter in the quasi-static regime and prevents it from supporting genuinely flat circular velocities.
These two outcomes point to a natural next step: to search for deformations that do break the hypersurface-deformation algebra—hence allowing controlled nonconservation—while yielding an effectively isotropic, dust-like source in the quasi-static regime, a combination that could plausibly evade the anisotropic obstruction and produce truly flat rotation curves.

By contrast, the HL branch exploits precisely what the GR deformation lacks:
the absence of a pointwise Hamiltonian constraint in the projectable theory.
The non-vanishing local Hamiltonian density is not imposed as a constraint;
rather, it is propagated by its off-shell Poisson bracket with the global
projectable Hamiltonian. For \(\lambda\neq 1\), this local-density bracket
contains the \(D_i\pi\) term, and after the leftover Hamiltonian density is
repackaged as an effective dust component, its propagation becomes a controlled
nonconservation equation for that dust. The two branches are therefore a matched pair: one reduced LTB sector, one question about the effective sources it can support, and one structural difference between the arms, namely whether the Hamiltonian constraint is imposed pointwise. The HL arm is classical and infrared throughout, with $\lambda$ a low-energy coupling constrained by galactic kinematics and the driven density generated locally on the quasi-static background rather than inherited from earlier cosmological evolution. In this setting the would be dust component is sourced by a controlled violation of local energy conservation along the preferred foliation, and the nonconservation term can drive a scaling contribution to the effective dust density. Building on earlier results, we identified a restricted family of LTB backgrounds for which the HL source term yields a positive scaling contribution compatible with the desired halo behavior, and we derived sharp constraints on the parameters controlling that family. This improves the phenomenological viability relative to the sign pathologies that can appear in more naive implementations, while making explicit how tightly the mechanism constrains the allowed functional form of the background evolution. Crucially, matching observed galactic rotation speeds implies a strong degeneracy: either the infrared theory sits extremely close to the GR limit, or one pushes the background exponent toward a near-critical value that simultaneously suppresses the HL source term toward zero, making it difficult for the HL driven dust production to dominate present day galaxies.
At the same time, our HL result should be read with care: what we have obtained is a consistent existence class for positive, scaling dust production in the passive (“test-tube”) sense, not a fully backreacted solution of the coupled HL field equations with baryons. In particular, we have not yet found the corresponding areal-radius solution and we have not incorporated the mutual backreaction between the generated dust component and the geometry. Establishing whether the power-law ansatz can arise as an actual solution—possibly as an attractor in some regime—and embedding it in a model that includes ordinary dust are therefore the natural next steps. Pursuing these directions would turn the present construction into a predictive framework capable of confronting both rotation curves and relativistic observables such as lensing within a single, self-consistent HL spacetime.

\section{Acknowledgments}

I would like to thank João Magueijo for the precious and helpful discussions. Many thanks to Haya AlMuhanna, Nishanth Kumar, Henry Price and Thomas Cheng for all the time spent thinking together.


\begin{thebibliography}{0}%
\makeatletter
\providecommand \@ifxundefined [1]{%
 \@ifx{#1\undefined}
}%
\providecommand \@ifnum [1]{%
 \ifnum #1\expandafter \@firstoftwo
 \else \expandafter \@secondoftwo
 \fi
}%
\providecommand \@ifx [1]{%
 \ifx #1\expandafter \@firstoftwo
 \else \expandafter \@secondoftwo
 \fi
}%
\providecommand \natexlab [1]{#1}%
\providecommand \enquote  [1]{``#1''}%
\providecommand \bibnamefont  [1]{#1}%
\providecommand \bibfnamefont [1]{#1}%
\providecommand \citenamefont [1]{#1}%
\providecommand \href@noop [0]{\@secondoftwo}%
\providecommand \href [0]{\begingroup \@sanitize@url \@href}%
\providecommand \@href[1]{\@@startlink{#1}\@@href}%
\providecommand \@@href[1]{\endgroup#1\@@endlink}%
\providecommand \@sanitize@url [0]{\catcode `\\12\catcode `\$12\catcode `\&12\catcode `\#12\catcode `\^12\catcode `\_12\catcode `\%12\relax}%
\providecommand \@@startlink[1]{}%
\providecommand \@@endlink[0]{}%
\providecommand \url  [0]{\begingroup\@sanitize@url \@url }%
\providecommand \@url [1]{\endgroup\@href {#1}{\urlprefix }}%
\providecommand \urlprefix  [0]{URL }%
\providecommand \Eprint [0]{\href }%
\providecommand \doibase [0]{http://dx.doi.org/}%
\providecommand \selectlanguage [0]{\@gobble}%
\providecommand \bibinfo  [0]{\@secondoftwo}%
\providecommand \bibfield  [0]{\@secondoftwo}%
\providecommand \translation [1]{[#1]}%
\providecommand \BibitemOpen [0]{}%
\providecommand \bibitemStop [0]{}%
\providecommand \bibitemNoStop [0]{.\EOS\space}%
\providecommand \EOS [0]{\spacefactor3000\relax}%
\providecommand \BibitemShut  [1]{\csname bibitem#1\endcsname}%
\let\auto@bib@innerbib\@empty
\end{thebibliography}%


\begin{thebibliography}{99}

\bibitem{Rubin:1980zd}
V.~C.~Rubin, W.~K.~Ford~Jr. and N.~Thonnard,
Astrophys. J. \textbf{238} (1980), 471-487
doi:10.1086/158003

\bibitem{Bosma:1981zz}
A.~Bosma,
Astron. J. \textbf{86} (1981), 1825-1846
doi:10.1086/113063

\bibitem{Sofue:2001rv}
Y.~Sofue and V.~Rubin,
Ann. Rev. Astron. Astrophys. \textbf{39} (2001), 137-174
doi:10.1146/annurev.astro.39.1.137
[arXiv:astro-ph/0010594 [astro-ph]].

\bibitem{Salucci:2019dm}
P.~Salucci,
Astron. Astrophys. Rev. \textbf{27} (2019), 2
doi:10.1007/s00159-018-0113-1
[arXiv:1811.08843 [astro-ph.GA]].

\bibitem{Wechsler:2018wpk}
R.~H.~Wechsler and J.~L.~Tinker,
Ann. Rev. Astron. Astrophys. \textbf{56} (2018), 435-487
doi:10.1146/annurev-astro-081817-051756
[arXiv:1804.03097 [astro-ph.GA]].

\bibitem{Bullock:2017xww}
J.~S.~Bullock and M.~Boylan-Kolchin,
Ann. Rev. Astron. Astrophys. \textbf{55} (2017), 343-387
doi:10.1146/annurev-astro-091916-055313
[arXiv:1707.04256 [astro-ph.CO]].

\bibitem{PDG:2024rpp}
S.~Navas \textit{et al.} [Particle Data Group],
Phys. Rev. D \textbf{110} (2024), 030001
doi:10.1103/PhysRevD.110.030001

\bibitem{Schumann:2019eaa}
M.~Schumann,
J. Phys. G \textbf{46} (2019), 103003
doi:10.1088/1361-6471/ab2ea5
[arXiv:1903.03026 [astro-ph.CO]].


\bibitem{Sanders:2002pf}
R.~H.~Sanders and S.~S.~McGaugh,
Ann. Rev. Astron. Astrophys. \textbf{40} (2002), 263-317
doi:10.1146/annurev.astro.40.060401.093923
[arXiv:astro-ph/0204521 [astro-ph]].

\bibitem{Ferreira:2009mm}
P.~G.~Ferreira and G.~D.~Starkman,
Science \textbf{326} (2009), 812-815
doi:10.1126/science.1172245
[arXiv:0911.1212 [astro-ph.CO]].

\bibitem{Bekenstein:2010mg}
J.~D.~Bekenstein,
in \textit{Particle Dark Matter: Observations, Models and Searches},
ed.~G.~Bertone (Cambridge Univ. Press, Cambridge, 2010) Chap.~6, pp.~99--118
doi:10.1017/CBO9780511770739.007
[arXiv:1001.3876 [astro-ph.CO]].




\bibitem{Milgrom:1983ca}
M.~Milgrom,
Astrophys. J. \textbf{270} (1983), 365-370
doi:10.1086/161130.

\bibitem{HL}
P.~Horava,
Phys. Rev. D \textbf{79}, 084008 (2009)
doi:10.1103/PhysRevD.79.084008
[arXiv:0901.3775 [hep-th]].

\bibitem{Magueijo:2024mach}
J.~Magueijo,
\textit{Phys.\ Lett.\ B} \textbf{858}, 139001 (2024)
doi:10.1016/j.physletb.2024.139001
[arXiv:2312.07597 [hep-th]].

\bibitem{Magueijo:2024dmssr}
J.~Magueijo,
\textit{Phys.\ Rev.\ D} \textbf{109}, 124026 (2024)
doi:10.1103/PhysRevD.109.124026
[arXiv:2404.15809 [hep-th]].

\bibitem{Magueijo:2024stsb}
J.~Magueijo,
\textit{Phys.\ Rev.\ D} \textbf{110}, 084050 (2024)
doi:10.1103/PhysRevD.110.084050
[arXiv:2406.17428 [gr-qc]].

\bibitem{Bassani:2025hmu}
P.~M.~Bassani and J.~Magueijo,
Phys.\ Rev.\ D \textbf{111}, 103529 (2025)
doi:10.1103/PhysRevD.111.103529
[arXiv:2502.00081 [gr-qc]].

\bibitem{Magueijo:2023eco}
J.~Magueijo,
Phys.\ Rev.\ D \textbf{108}, no.~10, 103514 (2023)
doi:10.1103/PhysRevD.108.103514
[arXiv:2306.08390 [hep-th]].

\bibitem{Carroll:2000fy}
S.~M.~Carroll,
Living Rev.\ Rel.\ \textbf{4} (2001), 1
doi:10.12942/lrr-2001-1
[arXiv:astro-ph/0004075 [astro-ph]].

\bibitem{shinji_review}
S.~Mukohyama,
Class. Quant. Grav. \textbf{27}, 223101 (2010)
doi:10.1088/0264-9381/27/22/223101
[arXiv:1007.5199 [hep-th]].

\bibitem{Bassani:2025hl}
P.~M.~Bassani, J.~Magueijo, and S.~Mukohyama,
\textit{J.\ Cosmol.\ Astropart.\ Phys.}\ \textbf{07} (2025) 032
doi:10.1088/1475-7516/2025/07/032
[arXiv:2408.03793 [gr-qc]].

\bibitem{Mukohyama:2009dm}
S.~Mukohyama,
Phys. Rev. D \textbf{80}, 064005 (2009)
doi:10.1103/PhysRevD.80.064005
[arXiv:0905.3563 [hep-th]].



\bibitem{Arnowitt:1959ah}
R.~Arnowitt, S.~Deser and C.~W.~Misner,
Phys.\ Rev.\ \textbf{116}, 1322--1330 (1959)
doi:10.1103/PhysRev.116.1322.

\bibitem{Arnowitt:1962dg}
R.~Arnowitt, S.~Deser and C.~W.~Misner,
in \textit{Gravitation: An Introduction to Current Research},
ed.~L.~Witten (John Wiley \& Sons, New York, 1962) Chap.~7, pp.~227--265.

\bibitem{Jha:2023hamadm}
R.~Jha,
SciPost Phys.\ Lect.\ Notes \textbf{73} (2023)
doi:10.21468/SciPostPhysLectNotes.73
[arXiv:2204.03537 [gr-qc]].

\bibitem{stephani}
H.~Stephani, D.~Kramer, M.~MacCallum, C.~Hoenselaers and E.~Herlt,
\textit{Exact Solutions of Einstein's Field Equations}, 2nd ed.
(Cambridge Univ.\ Press, Cambridge, 2003)
doi:10.1017/CBO9780511535185.

\bibitem{Enqvist:2007vb}
K.~Enqvist,
Gen. Rel. Grav. \textbf{40}, 451-466 (2008)
doi:10.1007/s10714-007-0553-9
[arXiv:0709.2044 [astro-ph]].

\bibitem{Goncalves:2001rv}
S.~M.~C.~V.~Goncalves,
Phys. Rev. D \textbf{63}, 124017 (2001)
doi:10.1103/PhysRevD.63.124017
[arXiv:gr-qc/0107087 [gr-qc]].

\bibitem{Vaz:2011cqsd}
C.~Vaz and L.~Witten,
Gen.\ Relativ.\ Gravit.\ \textbf{43}, 3429--3449 (2011)
doi:10.1007/s10714-011-1240-4
[arXiv:1111.6821 [gr-qc]].

\bibitem{Dirac}P.~Dirac, “Lectures on Quantum Mechanics”, Belfer Graduate School of Science, Yeshiva
University Press, New York, 1964.

\bibitem{DiracCanadian}P.~A.~M.~Dirac,
Can. J. Math. \textbf{2}, 129-148 (1950)
doi:10.4153/CJM-1950-012-1.

\bibitem{Thiemann}
T.~Thiemann,
\textit{Modern Canonical Quantum General Relativity},
Cambridge Monographs on Mathematical Physics
(Cambridge Univ.\ Press, Cambridge, 2007)
doi:10.1017/CBO9780511755682.

\bibitem{Bojowald:2011book}
M.~Bojowald,
\textit{Canonical Gravity and Applications: Cosmology, Black Holes, and Quantum Gravity}
(Cambridge Univ.\ Press, Cambridge, 2011)
doi:10.1017/CBO9780511921759.

\bibitem{Barriola:1989hx}
M.~Barriola and A.~Vilenkin,
\textit{Phys.\ Rev.\ Lett.}\ \textbf{63} (1989) 341
doi:10.1103/PhysRevLett.63.341.

\bibitem{Letelier:1979ej}
P.~S.~Letelier,
\textit{Phys.\ Rev.\ D}\ \textbf{20} (1979) 1294
doi:10.1103/PhysRevD.20.1294.


\bibitem{Curiel:2017primerEC}
E.~Curiel,
in \textit{Towards a Theory of Spacetime Theories},
ed.~D.~Lehmkuhl, G.~Schiemann and E.~Scholz
(Birkh\"auser, New York, 2017) pp.~43--104
doi:10.1007/978-1-4939-3210-8\_3
[arXiv:1405.0403 [physics.hist-ph]].


\bibitem{Treu:2010SLreview}
T.~Treu,
Annu.\ Rev.\ Astron.\ Astrophys.\ \textbf{48}, 87--125 (2010)
doi:10.1146/annurev-astro-081309-130924.

\bibitem{Komar:1963pde}
A.~Komar,
Phys.\ Rev.\ \textbf{129}, 1873--1876 (1963)
doi:10.1103/PhysRev.129.1873.

\bibitem{Sofue:2017rm}
Y.~Sofue,
Publ. Astron. Soc. Jpn. \textbf{69} (2017) no.1, R1
doi:10.1093/pasj/psw103
[arXiv:1608.08350 [astro-ph.GA]].

\bibitem{Persic:1995ru}
M.~Persic, P.~Salucci and F.~Stel,
Mon. Not. Roy. Astron. Soc. \textbf{281} (1996), 27-47
doi:10.1093/mnras/278.1.27
[arXiv:astro-ph/9506004 [astro-ph]].


\end{thebibliography}
\end{document}